\documentclass[preprint,pre,onecolumn]{revtex4}
\pdfoutput=1
\usepackage{hyperref}
\usepackage{graphicx}
\usepackage{amsmath}
\usepackage{amsmath}
\usepackage{amsfonts}
\usepackage{amssymb}
\usepackage{latexsym}
\usepackage{ifpdf}

\begin{document}

\title{Cooperation and its evolution in growing systems with cultural reproduction}

\begin{abstract}

We explore the evolution of cooperation in the framework of the evolutionary game theory using the prisoner's dilemma as metaphor of the problem. We present a minimal model taking into account the growing process of the systems and individuals with imitation capacity. We consider the topological structure and the evolution of strategies decoupled instead of a coevolutionary dynamic. We show conditions to build up a cooperative system with real topological structures for any natural selection intensity. When the system starts to grow, cooperation is unstable but becomes stable as soon as the system reaches a small core of cooperators whose size increase when the intensity of natural selection decreases. Thus, we reduce the emergence of cooperative systems with cultural reproduction to justify a small initial cooperative structure that we call \emph{cooperative seed}. Otherwise, given that the system grows principally as cooperator whose cooperators inhabit the most linked parts of the system, the benefit-cost ratio required for cooperation evolve is drastically reduced compared to the found in static networks. In this way, we show that in systems whose individuals have imitation capacity the growing process is essential for the evolution of cooperation.

\end{abstract}

\author{Ignacio Gomez Portillo}
\affiliation{Grup de F\'isica Estad\'istica, Departament de F\'isica, Universitat Aut\'onoma de Barcelona, 08193 Barcelona, Spain}
\maketitle

\section{Introduction}

\bigskip

What is cooperation? By definition, cooperation is the process of acting or working together. Some examples of systems where cooperation is present are the cells working together in order to form multicellular organisms and multicellular organisms working together to form societies such as anthills and the human society. It is noteworthy that cooperating biological systems make up new organisms that in turn can cooperate to form a new one with higher complexity. This shows that cooperation is a mechanism through which nature increases complexity of life. In this way, if there are organisms made up by other cooperative organisms, what would happen when for some reason these stop cooperating? What would happen to us if our cells stop cooperating? We easily note that the lack of cooperation means the death of the most complex organism. Therefore, understanding the mechanisms that allow cooperation to emerge and prevail is essential.

\bigskip

Cooperation \cite{G1,G2,G3,G4} is ubiquitous in biological systems and it supposes an evolutionary transition from isolated individuals to groups. Understanding how this transition arises within the framework of Darwinian theory is a big conceptual challenge that has received a lot of attention in the last fifty years \cite{G5,G6,G7,G8,G9,G10,G11,G12,G13,G14}. In this long and successful tradition, the first formal attempt to solve the problem was introduced by Hamilton in his celebrated paper of 1964 \cite{G5}. In this work cooperation is favored by natural selection if the genetic relatedness between individuals exceeds the benefit-cost ratio of the altruistic act.

\bigskip

The cooperation problem is usually studied in the framework of the evolutionary game theory \cite{G15,G16}. In this, the system can be represented by a network whose nodes are the individuals and the links represent who interacts with whom. Usually, the interactions are modeled by the Prisoner's Dilemma where each individual takes a strategy and plays against its neighbors. There are two strategies: cooperation ($C$) or defection ($D$). A cooperator gives a benefit $b$ and incurs in a cost $c$ for each individual it interacts with, where $b>c$ is required to define the game correctly. In contrast, a defector does not give benefits and has no costs, but it can receive the benefit $b$ from those cooperators it interacts with. The result of interactions of each individual defines its payoff $P$. The system evolves replicating more often successful strategies. Therefore, natural selection is introduced through payoff since individuals with a higher one reproduce more. The reproduction can be either genetic or cultural. When it is genetic, the most successful individuals have the most offspring. When reproduction is cultural the strategies of successful individuals are more likely to be imitated by their neighbors. It is worth noting that a case of cultural reproduction is indistinguishable from a death-bird process \cite{N16} when just the strategy of the individual is taken into account. However, throughout the text we will refer only to cultural reproduction considering that individuals have imitation capacity.

\bigskip

Until the early nineties the cooperation problem was only studied in fully connected systems in which unilateral defection is always the best strategy. Therefore defection is favoured by natural selection and invades the whole system. In order to overcome this problem, it was necessary to introduce other features to the system increasing the strategic complexity. In 1981, Axelrod and Hamilton. \cite{G6} showed that if the probability of new encounters between individuals is high enough, cooperation based on reciprocity can evolve. Furthermore, other features such as reputation \cite{G10} to individuals or punishment \cite{G14} for defectors were introduced.

\bigskip

However, it is well known that real systems are far from being fully connected. In a pioneer paper \cite{N1}, Nowak and May considered the problem over two-dimensional regular lattices showing that if individuals only have information of a neighborhood instead of the whole system, there are conditions in which cooperation prevail without the need to increase strategic complexity. Recently, together with the development of the graph theory, the cooperation problem has been studied considering more realistic topological structures of interactions \cite{N2,N3,N4,N5,N6,N7,N8,N9,N10,N11,N12,N13,N14,N15}. In particular, Santos \emph{et al.} \cite{N3,N4} showed how important the degree heterogeneity of the network could be for cooperation. In these works \cite{N1,N2,N3,N4,N5,N6,N7,N8,N9,N10,N11,N12,N13,N14,N15} the problem is considered assuming static networks large enough to ensure that the topological features under study are well developed. Initially each individual adopts cooperation o defection with equal probability, in order to not favor any strategy. After this, the system evolves updating strategies following an evolutionary game. Even though it has been shown that some topological features favor the sustainability of cooperation, the problem has not been completely solved given the high benefit-cost ratio required to promote cooperation with respect to those observed in nature. In particular, the difficulty present in static systems becomes evident when the system increase the average number of links \cite{N16,N5}. In particular, static systems present the following two limitations.

\bigskip

On the one hand, in many real systems the links are not static but there is a rewiring process. This has motivated the development of mechanisms with coevolution between the topological structure and the evolutionary game \cite{C1,C2,C3,C4,C5,C6}. In this way, the result of the game changes the topology which in turn changes the game result and so on until the cooperation level and the topological properties become time independent. These coevolutionary dynamics have shown to be a powerful mechanism to support cooperation as well as to produce the emergence of topologies of real systems. Furthermore, it is important to note that these models also take into account systems with degree heterogeneity and individuals with only local information. These features seem to be essential ingredients for cooperation.

\bigskip

On the other hand, cooperative biological systems do not emerge spontaneously fully developed but they are formed by a growing process from one or few individuals. Besides, it is remarkable that in all stages the system has levels of cooperation higher than half of the population. Another essential feature of biological systems is the possibility of mutation which in the context of cooperation means a spontaneous change of strategy. Therefore, taking into account the growing process and the possibility of mutations, we formulated the cooperation problem as follows:

\bigskip

- What conditions should the system have to grow as a highly cooperative system?

- What are the system properties that allow to resist the emergence of mutant defectors in a cooperative system?

\bigskip

Here we explore the cooperation problem considering the growing process of the system and individuals with imitation capacity. It is noteworthy that the growing process has also been considered \cite{C5} as a mechanism to introduce co-evolutionary dynamics. We consider that the topological structure and the evolutionary game are decoupled. We focus our attention in the first question assuming that the probability of mutation is low enough to be neglected, this is a routinely simplification in literature \cite{N16,N1,N2,N3,N4,N5,N6,N7,N8,N9,N10,N11,N12,N13,N14,N15}. Nevertheless, it is remarkable that this alternative way of approaching the problem is general, since it is applicable to all cooperative system in which there are mutations and it has been generated by network growth. Otherwise, the second question could be useful for cancer research given the similarity between defectors and tumor cells.

\bigskip

\section{The model}

The system is represented by a network of interconnected nodes. Each node is an individual who interacts with those who are connected with it. We considered all nodes equal and all connections undirected and with equal weight. Each interaction between individuals is modeled by a round of the prisoner's dilemma defined in the introduction. This definition allows to reduce the game parameters to one defined by the benefit-cost ratio $r=b/c$. Each individual only plays against all its neighbors. If the individual $i$ is a cooperator and interacts with $k_{i}$ neighbors, it receives a payoff $P_{i}=bk_{i}^{c}-ck_{i}$, where $k_{i}^{c}$ is the number of neighbors who are cooperators. When $i$ is a defector connected to $k_{i}$ neighbors, it receives a payoff $P_{i}=bk_{i}^{c}$. Here, we consider a process based on pairwise comparison between individuals to introduce cultural reproduction. We perform the strategy updates in a synchronous way, \emph{i.e.} all nodes are updated simultaneously. A complete update of the system is called a generation. When node $i$ updates strategy, it takes the strategy of a random chosen neighbor with probability $w$, which is function of $P_j - P_i$. On the other hand, the node $i$ keeps its strategy with probability $1-w$. Since natural selection favor successful strategies, $w$ increase with $P_j - P_i$. We define $w=w(P_j - P_i)$ as \cite{C5,C6,F2,F3,F4}, choosing the Fermi function from statistical physics, which is given by

\bigskip

\begin{equation}
w_{j\rightarrow i}=\frac{1}{1+e^{-\beta (P_{j}-P_{i})}}\text{ ,}
\end{equation}

The parameter $\beta$, which corresponds to an inverse temperature in statistical physics, controls the intensity of natural selection. Small $\beta$ (high temperature) means that selection is almost neutral, whereas for large $\beta$ (low temperature) selection can become arbitrarily strong. With decreasing intensity of selection $\beta$, the probability for imitation of the advantageous strategy in the population decreases from $1$ to $1/2$, selection becoming neutral for $\beta=0$. Moreover, it is noteworthy that this update rule allows the node $i$ to take the strategy of node $j$ even if $P_{i}>P_{j}$. This is important because introduce errors in individuals decision. Also, it is important to state that many other kinds of update rules \cite{F5,F6} can be taking into account. Furthermore, the results cannot be extended directly to any update rule since they usually are dependent on this choice \cite{F6}.

\bigskip

Actual cooperative systems grow from one or a few individuals by incorporating new ones through two different mechanisms. On the one hand, the new individuals may come from the reproduction of individuals of the system and, therefore, related genetically. On the other hand, the system grows by the incorporation of independent individuals who could not be related genetically to existing individuals. However, unless otherwise specified, the proportion of new individuals coming from each of the two sources does not change the results. The system is built up starting from $N_{0}$ arbitrarily connected nodes. Then, the system grows by adding new individuals with $L=N_{0}$ links. We studied two ways in which the $L$ links are attached to the system: random and preferential attachment \cite{N17}. In the latter case, new nodes are connected with a probability proportional to the degree of already existing ones, which produces a power-law degree distribution $P(k)\sim k^{-\gamma }$ with exponent $\gamma \simeq 2,9$ in the thermodynamic limit $N \rightarrow \infty$. When the nodes are randomly connected, the system acquires a degree distribution $P(k)$ that decays exponentially. We choose these growing mechanisms to cover a wide range of degree heterogeneities.

\bigskip

To take into account the growing process and the strategies update simultaneously it is necessary to define a temporal scale. We accomplished this through the system size assuming that $\frac{dN(t)}{dt}=aN(t)$, where $a$ is the growth rate. When the elimination of nodes is not considered, the growth rate is equal to the frequency with which individuals are incorporated to the system. We consider that $a$ is constant over time. Therefore, the population grows exponentially as $N(t)=N(t_0)e^{a(t-t_0)}$. Between two generations the system grows a constant time $\Delta t$. So if a strategy update is performed when the system has a population $N(t_0)$, then the next one is performed when the system has a population $N(t_0+\Delta t)=N(t_0)e^{a\Delta t}=N(t_0)(1+n)$, where $n$ is time independent for exponential growth. Consequently, the strategies are updated each time the system grows a fraction $n$, it is important to note that $a\Delta t\simeq n$ when $a\Delta t<<1$, which give a connection between $n$ and the measurable value $a$. Also, the model can be extended to consider other kinds of growth as for instance the logistic one, where the system initially grows exponentially in time but then it slows down until the system acquires a maximum size called carrying capacity of the system. Under this kind of behavior, it is expected that the frequency with which new individuals are incorporated to the system decreases in time and, therefore, the growing proportion $n$ before a strategy update also decrease. It has been checked that the results do not change significantly for values ​​of $n$ less than those shown. Therefore, the conclusions reached through the paper do not change when it is performed this kind of consideration. In particular, if the growth rate decreases i.e. $n$ tends to zero, as we shall show, the level of cooperation increase into the system when conditions allow cooperation to survive.

\bigskip

Finally, we must define the way individuals choose their first strategy. The Darwinian Theory restricts how strategies evolve but it does not carry information about the first strategy that each individual takes  in its first interaction. In the literature\cite{N1,N2,N3,N4,N5,N6,N7,N8,N9,N10,N11,N12,N13,N14,N15}, cooperation is the first strategy taken by individuals with probability $P_{c}=1/2$, assuming no bias towards any strategy.  For our purposes we can relax this condition to just $P_{c}\neq 0$ in order to introduce cooperation into the system. 

\bigskip

\section{Results}

\bigskip

The topological structures that emerges from the growing process, as the degree distribution, are macroscopic features of the system. Therefore, when the system begins to grow these topological structures are weakly developed due to lack of statistical \cite{N18}. As stated in the introduction, the macroscopic topological features play an important role in the cooperation problem, so it is expected that the proposed model behaves differently when the system starts to grow and when it reaches well defined topological structures. In this way, we divide the model into two parts in order to explore the model numerically and to perform a clear analysis from the literature \cite{N16,N1,N2,N3,N4,N5}. First, we assume that there are conditions under which the system reaches a state of $N_i$ cooperators. We take $N_i$ large enough to ensure the degree distribution well developed. From this structure, we seek the conditions required to maintain a stable and high level of cooperation when the system grows by the incorporation of defectors. This simulates the worst condition that a cooperative system must resist. Second, using the conditions found, we look for the minimum $N_i=N_c$ of cooperators that ensure the stability of the cooperation level within the system.

\bigskip

\begin{figure}[!hbt] \centering
\includegraphics[angle=0,scale=0.7]{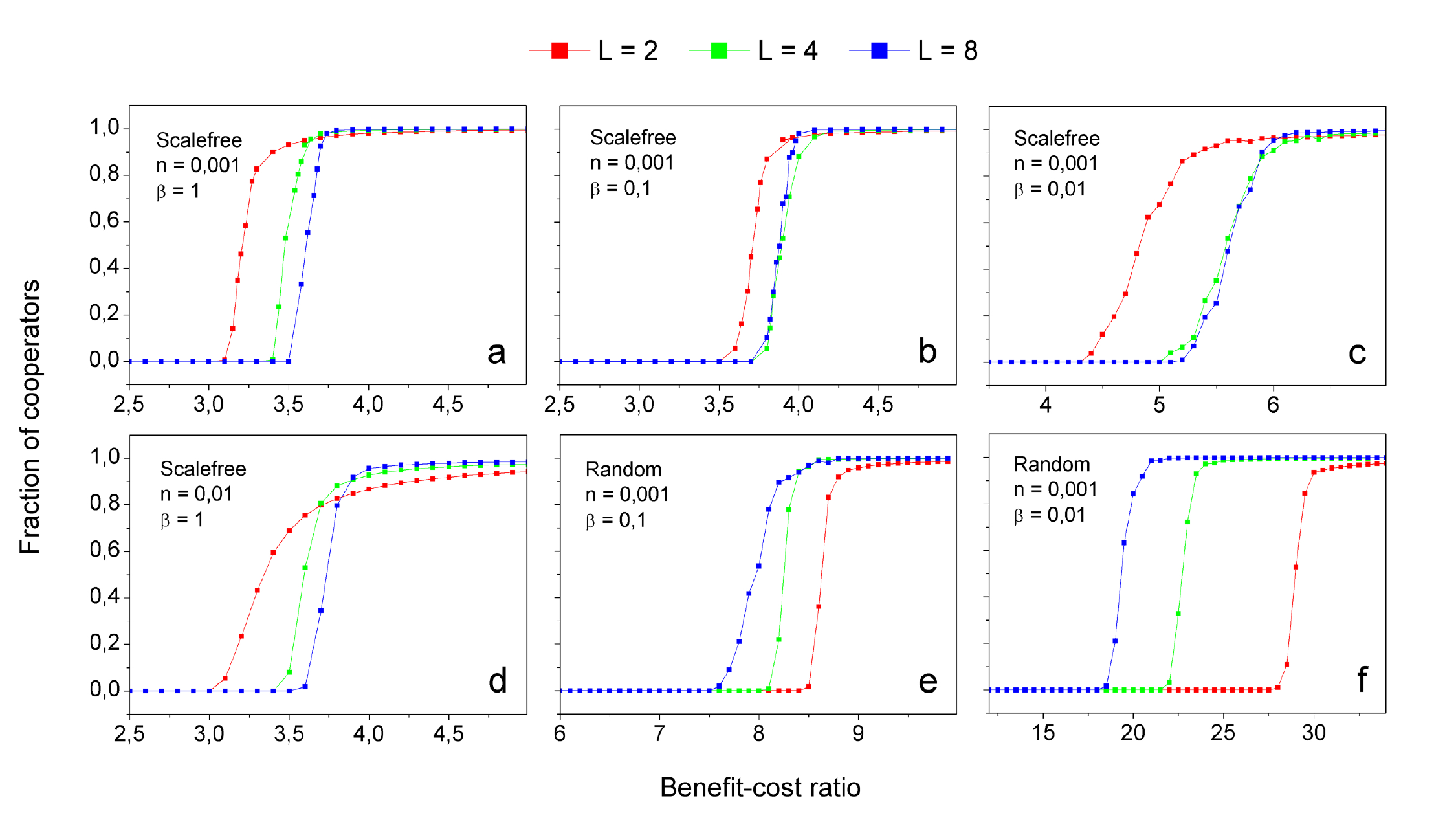}
\caption{\textbf{Conditions to maintain cooperation into the system for scale-free and random networks.}{\protect\footnotesize \ }Fraction of cooperators as a function of the benefit-cost ratio for different parameter values. Each figure presents a phase transition whose critical benefit-cost ratio $r_c$ depends mainly on the degree heterogeneity of the system and the intensity of natural selection. A high degree heterogeneity reduces the critical benefit-cost ratio value. Moreover, the critical value decreases when natural selection increases.}
\end{figure}

\bigskip

In Fig. 1 we show the numerical results for the fraction of cooperators into the system as a function of the benefit-cost ratio $b/c$ for different parameter values of the model and for the two proposed growing mechanisms. For these results we consider $N_{i}=5000$. We will later show that with a much smaller structure is sufficient. We let the system evolve until a stationary regime is reached. The stationary regime is characterized by a stable average level of cooperation $<c>$ which is the fraction of cooperators in the system. In order to make an extensive sampling of networks realizations we have performed $100$\ numerical simulations for each set of values of parameters studied and averaged accordingly to the value\ $<c>$ found in each realization. Also, it is important to note that the system does not evolve as a Moran process \cite{M1} because the system grows in time instead of having a fixed population. Therefore, when $0<P_{c}<1$ the system cannot reach permanently the absorbing states $<c>=1$ or $<c>=0$ usually found in Moran processes. Given that the system growth from $N_i$ cooperator by incorporation of defectors ($P_{c}=0$ from $N>N_i$), the systems can be absorbed by the state $<c>=0$ but not by the state $<c>=1$ for any value of $\beta$ and $r$.

\bigskip

In all figures we can see a phase transition from a noncooperative to a cooperative state when $r$ exceeds the critical threshold $r_c$. When the benefit-cost ratio is above $r_c$, with $n$ and $\beta $ fixed, the cooperative system resists the incorporation of new defectors reaching a nonzero and stable value of $<c>$ into the system. But, if the $b/c$ is below the threshold, the new defectors are strong enough to invade the whole system by imitation. When $r>r_c$ the systems is almost fully cooperative and defectors inhabit the less linked nodes. Under these circumstances the payoff $P$ is almost proportional to the degree $k$ of the nodes and the most connected ones are cooperators connected mainly to other cooperators. Thus, if $b/c$ is high enough, new defectors may have a higher payoff $P$ than nodes with few links, but nodes with a larger number of links continue to have a higher one. Therefore, defection can invade part of the system but it is stopped as soon as the invasion reaches sufficiently linked nodes. After that, the system finds the way to restore cooperation. This is possible because, while defection invades the system, the payoff of defectors decreases because they reduce the number of neighbor cooperators, making them weaker. If for some reason $b/c$ decreases below the threshold value; defection begins to invade the system. But, if the invasion does not reach the most connected nodes, increasing $b/c$ above its critical value cooperation is restored into the system. Also, it is important to note that $r_c$ is independent of the system size $N$ when $N>N_i$.

\bigskip

In addition, as shown Fig. $1$, the critical value is not strongly dependent of $L$ which it is clear for scale-free networks. In particular, we have verified this behavior to $L=32$ without significant changes with the results shown. Therefore, considering that the mean degree $<k>\simeq 2L$ when $N>>L$, we may approximate $r_{c}$ as independent of $<k>$ when the system grows following the proposed models of network generation. This can be understood by approaching $<c>$ by $1$ when $r>r_{c}$. Under this condition the new defectors receive a Payoff proportional to $L$ given that all its neighbors are cooperators. Furthermore, we can assume that the new defector exploits cooperators with an average Payoff proportional to $<k>$. Therefore, given that the relation $<k>/L\simeq 2$ is independent of $L$, it is expected $r_{c}$ to be independent of $<k>$. It is important to note as it has been shown previously \cite{N16,N5,N8} that the conditions required for cooperation to evolve increase significantly with $<k>$ in static systems. 

\bigskip

As can be seen, for any $n$ and $\beta$ fixed, scale-free networks reduce drastically the critical threshold $r_c$ with respect to random networks. The probability that an individual with low degree it is linked with a highly connected one increases with the degree heterogeneity. When $r>r_c$, the most connected nodes are occupied by cooperators and the lowest ones by defectors. Therefore, higher degree heterogeneity generates an increasing number of strong cooperators that enhance payoff differences with defectors. Thus, the critical $r_c$ required to maintain cooperation into the system decreases. This shows again the importance of the degree heterogeneity on the sustainability of cooperation \cite{N3,N4,N5}. However, here $r_c$ is drastically reduced compared to those found in static systems. Besides, taking into account that the difference between $r_c$ obtained by static systems and growing ones increases with $<k>$, we conclude that the growing process is essential for the evolution of cooperation in systems whose individual have imitation capacity. The origin of this difference is determined by two fundamental characteristics caused by the growing process. On the one hand, the system always has high levels of cooperation instead of initially half of the population considered in static systems. On the other hand, cooperators inhabit the most linked parts of the system rather than being initially randomly distributed. This reduces the exploitation capacity of defectors to a minimum because it is proportional to its degree $k$.

\bigskip

Moreover, comparing figures $1a$ and $1d$ it can be observed that increasing the frequency of strategy updates (decreasing $n$) enhances the fraction of cooperators for a benefit-cost ratio $b/c$ over the critical value, although it is not drastically modified. Here, it is important to state that equivalent results can be obtained when $n$ decreases from the shown values. When $n\rightarrow 0$ the system becomes fully cooperative after each strategy update if benefit-cost ratio is over the critical value. When $n$ grows, the system size required to overcome the transient becomes bigger and this makes it computationally hard to study for $n$ higher than those shown. However, when so many defectors are incorporated into the system before a strategy update, the defectors that become cooperators for each strategy update are less than the defectors incorporated making it impossible to find conditions where cooperation spreads into the system in the long run. Also, we can observe that the critical value increases when natural selection becomes weaker ($\beta \rightarrow 0$). However, it is possible to find conditions that promote cooperation when the system grows for any natural selection intensity.

\bigskip

As we have shown, new defectors can invade a part of the system. Therefore, when the system is small these invasions can be enough to invade the whole system. Moreover,  in the early stages of the system and by the growing mechanisms used, the new individuals have a number of links similar the existing nodes, which makes the invasions produced by new defectors more difficult to stop. This shows that cooperation is unstable when the system begins to grow. To explore this behavior, we study the cooperation fixation probability $P_{f}$ as a function of the number $N_{i}$\ of initial cooperators. The cooperation fixation probability is defined as the probability that a system of $N_i$ cooperators continues being cooperative when it grows by the incorporation of defectors for a fixed benefit-cost ratio. To obtain $P_{f}$ we select a value $r$ just over the critical value $r_c$ and we perform $M$ simulations starting from a system of $N_{i}$ cooperators. Then, we compute the number $M_{c}$ of systems that reach $N=10^{4}$ with a fraction of cooperators $<c>$ bigger than $1/2$. Finally, we compute $P_{f}$ for each $N_{i}$ as $M_{c}/M$. For the result we take $M=500$.

\bigskip

In Fig. $2$ we show the cooperation fixation probability as a function of the number of initial cooperators $N_i$. The results shown have been obtained by preferential attachment growth. As it can be seen in Fig. $2a$, the cooperation fixation probability $P_{f}$ increases more slowly for lower values of $\beta $. However, in all cases $P_{f}$ increases rapidly reaching the value $1$ for some small system size $N_{c}$ that we call \emph{cooperative seed}. When the system reaches the $N_{c}$ individuals the \emph{cooperative seed} is ready and cooperation becomes stable when the system grows. Also, as can be observed in Fig 2b  the \emph{cooperative seed} is larger for larger $L$. Given the results shown in Figure 1 we can infer that the system will reach a size beyond which cooperation will be stable for the parameters and growing mechanism that are not shown. It is important to state that a similar behavior is found in static systems \cite{N3,N4,N5} since cooperation just can evolve from some system size.
\bigskip

\begin{figure}[!hbt] \centering
\includegraphics[angle=0,scale=0.65]{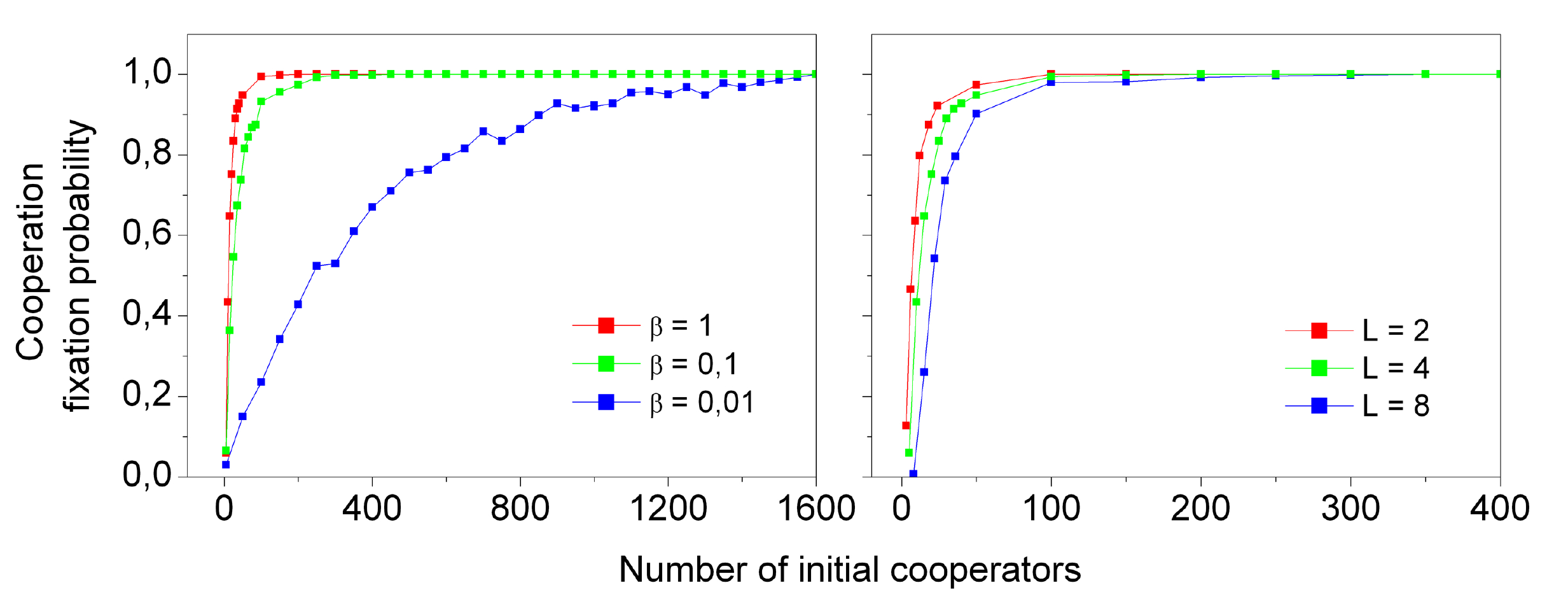}
\caption{\textbf{Cooperative seed.} Cooperation fixation probability $P_{f}$ as a function of number of initial cooperators $N_{i}$ for different parameter values when the system grows with preferential attachment. As can be seen, $P_{f}$ becomes one beyond some critical size $N_{c}$. From this structure cooperation is stable when the system grows. a) For these results we used the following values: $L=4$ and $n=0,001$. b) Here we use $\protect\beta =1$ and $n=0,001$.}
\end{figure}

\bigskip

It is important to note that the update strategy can be introduced in an asynchronous way without significant changes in the results. We use synchronous updates only for computational efficiency. Moreover, if instead of the Fermi rule we use the analog of replicator dynamics for finite populations \cite{N3,N4,N5} in a synchronous or asynchronous way it is obtained similar results to those shown with strong natural selection. We use the Fermi rule in order to tune the natural selection intensity.

\bigskip

Finally, the growing mechanisms presented have some unrealistic features that can worsen the conditions to maintain cooperation not only when the system is big, but mainly when it is small. First, it would be important to take into account that if new individuals are the result of genetic reproduction, the newborn have less intrinsic adaptability ($IA$) to the environment than their parents, where $IA$ is the adaptability of the individual living alone. This feature allows cooperation to be a better strategy than defection even when both have equal number of links. Also, it should be considered that newborns are not prepared to reproduce initially because they are not able to do it at early stages. This reduces the number of defector-defector links which weakens defection. Furthermore, it would be important to consider that the links of the new individuals are not performed simultaneously but one at a time. This is important because it gives the possibility to update the strategies when defectors have few links and therefore reducing their capacity to exploit cooperators because they reduce their payoff increasing probability to change strategy. Also, we assume $b/c$ constant for any system size, but this may not be a good approximation if the resources are limited; this could be particularly important when the system is small.

\bigskip

\section{Conclusions}

\bigskip

First, we have presented a simple model for the evolution of cooperation in growing systems whose individuals have imitation capacity. First, we have shown the existence of a critical benefit-ratio $r_c$ beyond which a cooperative system, with a well-defined degree distribution, maintain high level of cooperation when it grows by new defectors. If the frequency of strategy updates is high enough, there are a $r_c$ beyond which cooperation prevails into the system for the two growing mechanism explored and for any natural selection intensity. However, we have to bear in mind that conditions are improved when natural selection becomes stronger. We show for the considered growing mechanism that $r_c$ can be considered constant with the average degree $<k>$, this is an unprecedented result in structured populations. Besides, we have proved that $r_c$ is reduced when degree heterogeneity increase. This agrees with previous results \cite{N3,N4,N5}. However, the growing process reduces drastically $r_c$ required to cooperation evolve.

\bigskip

Secondly, we have studied the model when the system is small. We found that under this condition cooperation can disappear but it is increasingly less likely as the number of initial cooperators increases. Cooperation becomes stable beyond a certain size of initial cooperators, what we call \emph{cooperative seed}. This initial structure can be justified combining two features of the model: the nonzero probability in which individuals take cooperation as the first strategy and the cooperation fixation probability which grows fast when the initial cooperative structure becomes bigger. Clearly, this does not completely solve the problem of the emergence of cooperation since there are many cases in nature in which the emergence of cooperation is highly probable from the starting stages of the system. However, it is important to note that the model allows reduce the problem of the evolution of cooperation without mutations to justify the formation of the \emph{cooperative seed}. 

\bigskip

The principal features that allow the model to form a cooperative system are the growing process and the fact that new individuals introduce few links with respect to old ones. These two conditions ensure degree heterogeneity and individuals with local information that have been recognized as essential for cooperation. But the growing process also incorporates a new feature that was not previously taken into account improving the conditions required in previous works \cite{N1,N2,N3,N4,N5,N6,N7,N8}; now the system grows mainly formed by cooperators that dwell in the most connected part of the system.

\bigskip

We believe that, given the generality and simplicity of the model, it would be of great interest to test the validity of the results experimentally. We think it is interesting the way in which we have formulated the problem of cooperation in the introduction. An important consequence is that it allows explore the emergence of mutants assuming that the whole system is made up by cooperators. This could change the way the cooperation problem is addressed. Besides of looking for mechanisms to allow cooperation to invade the system, it would be interesting to search for mechanisms in which mutants defectors cannot invade a cooperative system. This new point of view allows to reinterpret some previous results \cite{N1,N2,N3,N4,N5,N6,N7,N8,N9,N10,N11,N12,N13,N14,N15} showing that even when a random fifty percent of a population of cooperators becomes defector there are conditions and topologies that restore cooperation into the system.

\bigskip

\section{\protect\bigskip Acknowledgements}

I acknowledge financial support from Generalitat de Catalunya under grant 2009-SGR-00164 and PhD fellowship 2011FIB787. I am grateful to the Statistical Physics Group of the Autonomous University of Barcelona. In particular, I would like to thank David Jou, Ninfa Radicella and Vicen\c c M\'{e}ndez for their unconditional support and assistance with the manuscript. To Ruben Requejo for introducing me to the fascinating world of cooperation. I also would to thank to Carles Calero, Daniel Campos, Eduardo Barba, Ezequiel Manavela Chiapero, Fernando Arizmendi, Ivan Duran Sancho, Nestor Ghenzi and Pablo M. Gleiser for valuable comments and help on the writing. To Maxi San Miguel and Yamir Moreno for the selfless help and enriching comments. To Silvia Carrascosa for giving me moral support throughout this work. Also, I am deeply grateful to Lucas Gomez Portillo for showing me the importance of never giving up in adversity.


\bigskip

\begin{thebibliography}{99}

\bibitem{G1} E.O. Wilson, \emph{Sociobiology} (Harvard Univ. Press, Cambridge, Massachusetts, 1975).

\bibitem{G2} J. Maynard-Smith, E. Szathm\'{a}ry, \emph{The major transitions in evolution} (Oxford Univ. Press, Freeman, Oxford, 1995).

\bibitem{G3} R.E. Michod, \emph{Darwinian dynamics: Evolutionary transitions in Fitness and individuality} (Princeton Univ. Press, Princeton, NJ, 1999).

\bibitem{G4} R. Trivers, Q. Rev. Biol. \textbf{46}, 35 (1971).

\bibitem{G5} W.D. Hamilton, J. Theor. Biol. \textbf{7}, 1 (1964).

\bibitem{G6} R. Axelrod, W.D. Hamilton, Science \textbf{211}, 1390 (1981).

\bibitem{G7} M.A. Nowak, Science \textbf{314}, 1560 (2006).

\bibitem{G8} M.A. Nowak and K. Sigmund, Nature \textbf{364}, 56 (1993).

\bibitem{G9} R.L. Riolo, M.D. Cohen, and R. Axelrod, Nature \textbf{414}, 441 (2001).

\bibitem{G10} M.A. Nowak and K. Sigmund, Nature \textbf{437}, 1291 (2005).

\bibitem{G11} A. Arenas, J. Camacho, J.A. Cuesta, and R. Requejo, J. Theor. Biol. \textbf{279}, 113 (2011).

\bibitem{G12} J. Sanjay and K. Sandeep, Proc. Nac. Acad. Sci. \textbf{98}, 543 (2001).

\bibitem{G13} F.C. Santos, M.D. Santos, and J.M. Pacheco, Nature \textbf{454}, 212 (2008).

\bibitem{G14} E. Fehr and S. G\"{a}chter, Nature \textbf{415}, 137 (2002).

\bibitem{G15} J. Hofbauer and K. Sigmund, \emph{Evolutionary Games and Population Dynamics} (Cambridge University Press, Cambridge, England, 1998).

\bibitem{G16} H. Gintis, \emph{Game Theory Evolving} (Princeton University, Princeton, NJ, 2000).

\bibitem{N16} H. Ohtsuki, C. Hauert, E. Lieberman, and M.A. Nowak, Nature \textbf{441}, 502 (2006).

\bibitem{N1} M.A. Nowak and R.M. May, Nature \textbf{359}, 826 (1992).

\bibitem{N2} G. Abramson, M. Kuperman, Phys. Rev. E \textbf{63}, 030901 (2001).

\bibitem{N3} F.C. Santos and J.M. Pacheco, Phys. Rev. Lett. \textbf{95}, 098104 (2005).

\bibitem{N4} F.C. Santos, J.M. Pacheco, and T. Lenaerts, Proc. Nac. Acad. Sci. \textbf{103}, 3490  (2006).

\bibitem{N5} Y.-S Chen, H. Lin and, C.-X Wu, Physica A \textbf{385}, 379 (2006).

\bibitem{N6} S. Assenza, J. G\'{o}mez-Garde\~{n}es, and V. Latora, Phys. Rev. E \textbf{78}, 017101 (2008).

\bibitem{N7} J. G\'{o}mez-Garde\~{n}es, M. Campillo, L.M. Flor\'{\i}a and Y. Moreno, Phys. Rev. Lett. \textbf{98}, 108103 (2007).

\bibitem{N8} X. Chen, F. Fu, and L. Wang, Physica A \textbf{378}, 512 (2008).

\bibitem{N9} L. Luthi, E. Pestelacci, and M. Tomassini, Physica A \textbf{387}, 955 (2008).

\bibitem{N10} Y.-K. Liu, Z. Li, X.-J. Chen, and L. Wang, Chin. Phys. Lett. \textbf{26}, 048902 (2009).

\bibitem{N11} F. Fu, L. H. Liu, and L. Wang, Eur. Phys. J. B \textbf{56}, 367 (2007).

\bibitem{N12} F. Fu, X. Chen, L. Liu, and L. Wang, Phys. Lett. A \textbf{371}, 58 (2007).

\bibitem{N13} Z. -X. Wu, J. -Y. Guan, X. -J. Xu, and Y. -H. Wang, Physica A \textbf{379}, 672 (2007).

\bibitem{N14} N. Masuda, Proc R. Soc. B \textbf{274}, 1815 (2007).

\bibitem{N15} A. Szolnoki, M. Perc, and Z. Danku, Physica A \textbf{387}, 2075 (2008).

\bibitem{C1} J. M. Pacheco, A. Traulsen and M. A. Nowak, Phys. Rev. Lett. $\boldsymbol{97}$, 258103 (2006).

\bibitem{C2} F. C. Santos, J. M. Pacheco and T. Lenaerts, PLoS Comp. Biol. $\boldsymbol{2}$, 1284-1290 (2006).

\bibitem{C4} J. M. Pacheco, A. Traulsen, H. Ohtsuki and M. A. Nowak, J. Theor. Biol. $\boldsymbol{250}$, 723-731 (2007).

\bibitem{C3} J. Poncela, J. G\'{o}mez-Garde\~{n}es, L. M. Flor\'{\i}a, A. S\'{a}nchez and Y. Moreno, PLoS one $\boldsymbol{3}$, e2449 (2008).

\bibitem{C5} J. Poncela, J. G\'{o}mez-Garde\~{n}es, A. Traulsen and Y. Moreno, New J. of Phys. $\boldsymbol{11}$, 083031 (2009).

\bibitem{C6} M. Perc and A. Szolnoki, Biosystems $\boldsymbol{99}$, 109 (2010).

\bibitem{F2} A. Traulsen, J. M. Pacheco and M. A. Nowak, J. Theor. Biol. \textbf{246}, 522-529 (2007).

\bibitem{F3} G. Szab\'o, C. T\H{o}ke, Phys. Rev. E \textbf{58}, 69 (1998).

\bibitem{F4} C. Hauert, G Szab\'o, Am. J. Phys. \textbf{73}, 405 (2005).

\bibitem{F5} G. Szab\'o, G. F\'ath, Phys. Rep. \textbf{446}, 97 (2007).

\bibitem{F6} C. P. Roca, J. A. Cuesta, A. Sánchez, Phys. Life Rev. \textbf{6}, 208 (2009).

\bibitem{N17} A. L. Barab\'{a}si and R. Albert, Science \textbf{286}, 509 (1999).

\bibitem{N18} M. E. J. Newman, SIAM Review \textbf{45}, 167  (2003).

\bibitem{M1} T. Antal and I. Scheuring, Bull. Math. Biol. \textbf{68} 1923-1944 (2006).
 
\end{thebibliography}
\end{document}